\documentstyle[twocolumn,prb,aps,epsf]{revtex}

\begin{document}
\draft

\twocolumn[\hsize\textwidth\columnwidth\hsize\csname @twocolumnfalse\endcsname

\title{Vortex dynamics of a $d+is$-wave superconductor }

\author{Qunqing Li$^1$, Z. D. Wang$^1$, and Qiang-Hua Wang$^{1,2}$}

\address{$^1$Department of Physics, University of Hong Kong, Pokfulam Road,
Hong Kong, China}

\address{$^2$National Laboratory of Solid State Microstructures,
Institute for Solid State Physics,\\ Nanjing University, Nanjing 210093, China}

\maketitle

\begin{abstract}
The vortex dynamics of a $d+is$-wave superconductor is studied numerically
by simulating the time-dependent Ginzburg-Landau equations. The critical
fields, the free flux flow, and the flux flow in the presence of twin-boundaries
are discussed. The relaxation rate of the order parameter turns out to play
an important role in the flux flow. We also address briefly the intrinsic
Hall effect in $d$- and $d+is$-wave superconductors.
\end{abstract}


\vskip2pc
] 

\newpage 
In recent years the symmetry of the pairing function has been one of the
interesting topics in the field of high temperature superconductor. It is
widely accepted that the dominant pairing wave function has a $d_{x^2-y^2}$
symmetry, as supported by experiments using phase-sensitive devices, such as
Josephson junctions or Superconducting Quantum Interference Devices(SQUID).~
\cite{Wollman,Iguchi,Miller,Tsuei,Mathai} However, the subdominant pairing
channels such as $s$-wave or $d_{xy}$-wave channels are still possible.~\cite
{Sun,Ma,Ding} The mixed state of $s+d$ superconductors was first discussed
by Ruckenstein {\it et al.} ~\cite{Ruck} and that of $d+is$ superconductors
by Kotliar~\cite{Kot}. In Ref.[\cite{Kot}], it is pointed out that the
resonating-valence-bond mechanism can lead to $s$-wave and $d$-wave-like
Cooper pairings, and a mixture of $s$ and $d$ waves with a well-defined
relative phase close to $\theta=\pi/2$ is energetically favored.
Importantly, the superconducting state is time-reversal-symmetry (T
hereafter) breaking and the energy gap is nodeless unless $\theta=0$ or $\pi$%
. Moreover, surfaces and interfaces, grain boundaries, and other
pair-breaking defects have all been shown to enhance the T-breaking states.~
\cite{Kirt,Sig}

The vortex structure and vortex dynamics of $d$-wave superconductor have
been studied numerically in detail by several methods~\cite
{Soin,Xu,Berlin,Heeb,Wang}. Previous simulations for the $d+is$-wave
superconductors showed that the vortex structure of $d+is$-wave
superconductor is different from that of the $d$-wave superconductor~\cite
{Qqli}: The spatial profile of the moduli of $s$- and $d$-wave components in
the $d+is$-wave state has a two-fold symmetry, in contrast to the four-fold
symmetry of the magnetic-field-induced $s$-wave component of $d$-wave
superconductor. However, it turns out that such a $d+is$ state can only be
stabilized at extremely low temperatures. With increasing temperature, the
amplitude of $s$-wave component decreases and its symmetry changes from
two-fold to four-fold. Here we would like to extend our study to the vortex
dynamics of $d+is$ superconductors. For conventional superconductors the
free-flux-flow (FFF) resistance is linear in the magnetic field induction $B$
(with $B\ll B_{c2}$). The situation is not clear yet for high-$T_c$ and
other unconventional superconductors. Because of the multiple components of
the order parameter, this topic is highly nontrivial. We also study the
vortex motion in the presence of a twin-boundary and the intrinsic Hall
effect of $d+is$-wave superconductors.

We start with a model of an isotropic two-dimensional Fermi liquid with
attractive interactions in both $s$- and $d$- channels. Obviously, when only
one of the two interactions is attractive, the ground state is a pure state
with the appropriate pairing symmetry. When both channels are attractive,
the competition will lead to either a pure or a mixed pairing function. The
GL theory for a superconductor with two attractive channels has been
presented by Ren, Xu, and Ting~\cite{Ren} based on Gor'kov equations.
Assuming pure dissipative dynamics, the GL equations can be expressed as
follows: 
\begin{eqnarray}
0 &=&\left[ \eta _{s}\partial t-\alpha _{s}+\frac{4}{3}(|S|^{2}+|D|^{2})+%
{\bf \Pi }^{2}\right] S+  \nonumber \\
&&\frac{2}{3}D^{2}S^{\ast }+\frac{1}{2}({\bf \Pi }_{x}^{2}-{\bf \Pi }%
_{y}^{2})D  \label{Eq:S} \\
0 &=&\left[ \eta _{d}\partial t-\alpha _{d}+\frac{8}{3}|S|^{2}+|D|^{2}+{\bf %
\Pi }^{2}\right] D+  \nonumber \\
&&\frac{4}{3}S^{2}D^{\ast }+({\bf \Pi }_{x}^{2}-{\bf \Pi }_{y}^{2})S
\label{Eq:D} \\
0 &=&{\frac{\partial {\bf A}}{\partial t}+\kappa ^{2}({\bf \nabla }}\times 
{\bf \nabla }\times {\bf A}-{\bf \nabla }\times {\bf H_{e}})  \nonumber \\
&&+\left\{ S^{\ast }{\bf \Pi }S+\frac{1}{2}D^{\ast }{\bf \Pi }D+\frac{1}{2}%
\left[ S^{\ast }({\bf \Pi }_{x}-{\bf \Pi }_{y})D\right. \right.   \nonumber
\\
&&\left. \left. +D^{\ast }({\bf \Pi }_{x}-{\bf \Pi }_{y})S\right] +{\rm H.c}%
.\right\} {\rm .}  \label{Eq:A}
\end{eqnarray}
In these equations, the two order parameters, $S$ and $D$, are normalized by 
$\Delta _{0}=\sqrt{4/3\alpha \ln (T_{d}/T)}$ with $\alpha \approx 7\zeta
(3)/8(\pi T_{c})^{2}$, the space by the coherence length $\xi $, and the
vector potential ${\bf A}$ by $\Phi _{0}/2\pi \xi $ with $\Phi _{0}=h/2e$
being the flux quantum, respectively. In Eq.(1), $\alpha _{s}$ may be
expressed as a function of temperature $T$:~\cite{Ren,Note} 
\begin{equation}
\alpha _{s}=\ln (T_{s}/T)/\ln (T_{d}/T)  \label{Eq:As}
\end{equation}
where $T_{s}$ and $T_{d}$ may be viewed as the apparent superconducting
transition temperatures for the $s$-wave and the $d$-wave respectively, with 
$T_{s}\propto e^{-\frac{1}{N(0)V_{s}}}$ and $T_{d}\propto e^{-\frac{2}{%
N(0)V_{d}}}$. $\alpha _{d}=1$ unless specified otherwise. Here $N(0)$ is the
density of states at the Fermi surface, $V_{s}$ and $V_{d}$ are the
effective attractive interaction strengths in the $s$- and $d$-wave
channels, respectively. ${\bf \Pi }=i{\bf \nabla }+{\bf A}$, ${\bf \Pi }_{k}=%
\hat{x}_{k}\Pi _{k}$. $\kappa $ is the GL parameter. The time $t$ is
normalized by $\tau =\sigma _{n}\xi ^{2}$ with $\sigma _{n}$ the
normal-state conductivity of the superconductor. $\eta _{s,d}=\tau
_{s,d}/\tau $, with $\tau _{s,d}$ being the relaxation time of the $s$- and $%
d$- wave order parameters, respectively. For gapless conventional
superconductors, the relaxation time of the order parameter near $T_{c}$ is $%
\tau _{s}=\pi \hbar /8k(T_{c}-T)\propto \xi ^{2}$.~\cite{Gorkov} For
unconventional superconductors, it is plausible to assume the general
scaling law $\tau _{s,d}\sim \xi ^{z}$ with $z$ as the dynamical exponent,
which should hold at least near the critical temperature. In this case, it
is clear that $\eta _{s,d}\sim \xi ^{z-2}$ would depend on temperature if $%
z\neq 2$. However, in the absence of a microscopic theory on the relaxation
times, we shall content ourselves to treat them as phenomenological
parameters, the effect of which on the vortex dynamics is examined in this
paper.

The scale for the magnetic field is set by the upper critical field $H_{c2}$%
. Because of its nontrivial nature in the present system, we would like to
discuss it before moving to the dynamical properties of the vortices.
Following the same method described by Sigrist {\it et al.}~\cite{Sigr}, and
using the relation $\left[\Pi_{x},\Pi_{y}\right]=iB$, we introduce a pair of
operators $a=\frac{1}{\sqrt{2B}}(\Pi_{x}+i\Pi_{y})$, $a^{+}=\frac{1}{\sqrt{2B%
}}(\Pi_{x}-i\Pi_{y})$, and have the commutative relation $\left[a,a^{+}%
\right]=1$. Substituting $a$,$a^{+}$ into the linearized Eq.(\ref{Eq:S}) and
Eq.(\ref{Eq:D}) in the static case, we obtain: 
\begin{eqnarray}
& &(-2\alpha_s+2B(1+2\hat n))S+B(aa+a^{+}a^{+})D=0 ,  \nonumber \\
& &(-1+B(1+2\hat n)D+B(aa+a^{+}a^{+})S=0 ,  \label{Eq:Liner}
\end{eqnarray}
where $\hat {n}=a^{+}a$. Therefore, the order parameter can be expanded in
this occupation representation as $\Psi ={\sum\limits_{m=0}^{\infty}}\left (
\begin{array}{c}
s_{m}\mid m\rangle_{s} \\ 
d_{m}\mid m\rangle_{d}
\end{array}
\right )$, where $\hat {n}\mid m\rangle=m\mid m\rangle$. Eq.(\ref{Eq:Liner})
can not be solved exactly, so we treat it variationally by assuming $\Psi
=\left(
\begin{array}{c}
s_{0}\mid 0\rangle_{s} \\ 
d_0\mid 0\rangle_{d}
\end{array}
\right )+ \left(
\begin{array}{c}
s_2\mid 2\rangle_{s} \\ 
d_2\mid 2\rangle_{d}
\end{array}
\right ) + \left(
\begin{array}{c}
s_4\mid 4\rangle_{s} \\ 
d_4\mid 4\rangle_{d}
\end{array}
\right )$. $H_{c2}$ is determined by the condition that the ground state
energy of the eigenvalue problem Eq.(\ref{Eq:Liner}) is zero. It turns out
that the variational ground state is $\Psi=\left(
\begin{array}{c}
0 \\ 
d_0\mid 0\rangle_{d}
\end{array}
\right )  +\left(
\begin{array}{c}
s_2\mid 2\rangle_{s} \\ 
0
\end{array}
\right )  +\left(
\begin{array}{c}
0 \\ 
d_4\mid 4\rangle_{d}
\end{array}
\right )$, and the corresponding $H_{c2}$ is given by the root of the
equation 
\begin{equation}
60B^{3}-86B^2+9\alpha_{s}B^2+(10-10\alpha_{s})B+\alpha_{s}=0.  \label{Eq:Hc2}
\end{equation}
The temperature dependence of $H_{c2}$ calculated in this way is shown as
the solid line in Fig.(1). Here we use Eq.(\ref{Eq:As}) for the temperature
dependence of $\alpha_s$, and assume $T_{d}=100K$ and $T_{s}=90K$. In order
to check the reliability of the variational treatment, we also simulate the
upper critical field of such a $d+is$ superconductor numerically by solving
Eq.(\ref{Eq:S}) $\sim$ Eq.(\ref{Eq:A}) with the finite element method. In
practice, we fix the magnetic field induction $B$ by specifying one or two
vortices in a square unit cell with periodic boundary conditions. The side
length is varied so as to change the magnetic induction $B$. The technical
details of the simulation have been given else where.~\cite{Wang} The
external magnetic field $H$ can be derived from the Virial relation.~\cite
{Virial} With $H$ the Gibbs free energy density can be constructed, as shown
in Fig.2. We can read off $H_{c1}$ from the intersection of Meissner state
line and the mixed-state line, and $H_{c2}$ from the intersection of the
mixed-state line and the normal-state line. The simulation result for the
upper critical field is shown as black squares in Fig.1. The variational
result (solid line) appears to be in excellent agreement with the simulation
result, indicating that expanding the trial wave function up to the
occupation state of $|4\rangle$ is already rather reliable.~\cite{Wkim} Note
that $H_{c2}$ is larger than that of the pure $d$-wave superconductor, which
is always $B_0=\Phi_0/2\pi\xi^2$ in our case. From Fig.2, it is obvious that
the magnetic filed-induced transitions at $H=H_{c1}$ and $H=H_{c2}$ are both
of the usual second order transition, which should be compared to the
unusual first order transitions found numerically in some p-wave
superconductors.~\cite{Wang98}

\begin{figure}[btp]
\epsfxsize=8.5cm
\epsfbox{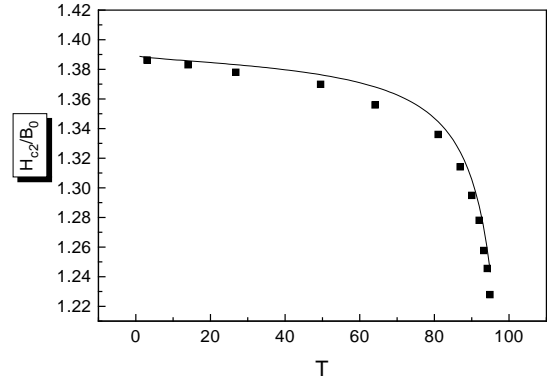}
\caption{The upper critical field $H_{c2}/B_0$ as a function of 
temperature $T$ with $B_0$ as the upper critical field of pure $d$-wave.
The solid line is the approximation result determined from Eq.(6). 
The scatter point is the numerical result. Here $T_s$ is set to $90K$, 
$T_d$ is set to $100K$.} 

\end{figure}

\begin{figure}[btp]
\epsfxsize=8.5cm
\epsfbox{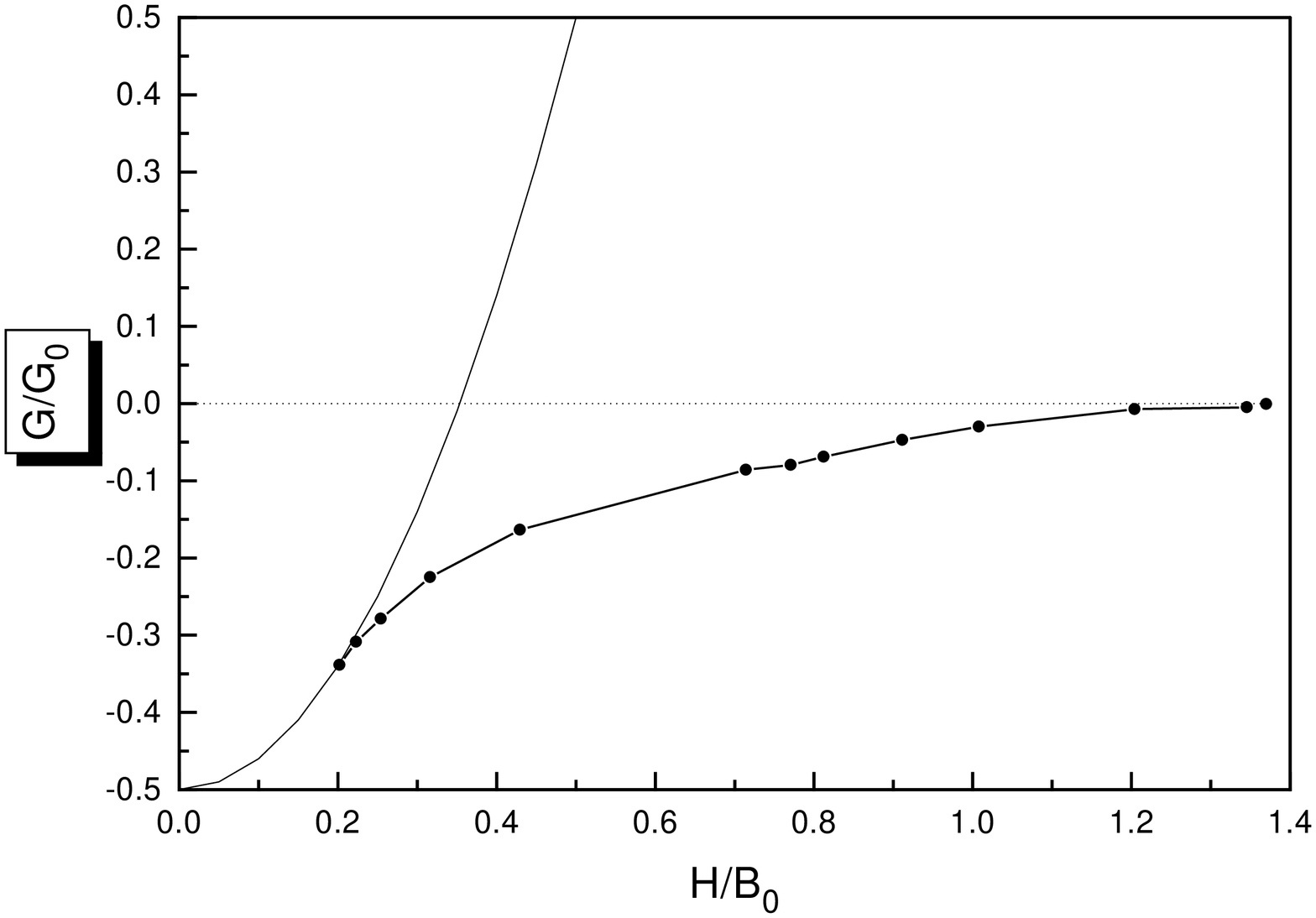}
\caption{Free vortex flow resistivity as a function of the applied field for 
$\eta_s=1$, $10$, and $0.01$. Here $\alpha_{s}=0.97$, $0.85$,
$0.67$, and $-1$ in (a), (b), (c), and (d) respectively.}
\end{figure}

We now discuss the vortex flow driven by a transport current ${\bf J}$ in
the mixed state (i.e., with $B<H_{c2}$). This is realized by by requiring $%
{\bf \nabla}\times {\bf H_{e}}={\bf J}$ in Eq.(\ref{Eq:A}). Here we have
chosen a gauge in which the electrostatic potential does not appear, so that
the local electric field is simply ${\bf E}=-\partial_{t}{\bf A}$. To
investigate the relaxation effects of the order parameter on the FFF
resistivity, we chose three values of the relaxation coefficient of the
order parameter $S$ and $D$ as :(1) $\eta_{s}=2\eta_{d}=1$, (2)$%
\eta_{s}=2\eta_{d}=10$ ($\gg 1$), (3) $\eta_{s}=2\eta_{d}=0.01$ ($\ll 1$).
The field dependence of the FFF resistivity is shown in Fig.3. Noticeably it
evolves from a convex to a concave with increasing $\eta_s$, albeit slight
difference exists at the four temperatures shown in the four panels. This is
the correct trend by general reasoning: The motion of vortices is equivalent
to the phase slipping of the order parameter,~\cite{Tinkham} so that a small
relaxation time means a quick rate of phase slipping, and thus a large
resistance. Thus, with increasing magnetic field the vortex flow resistivity
approaches the normal state resistivity more quickly. It follows from Fig.3
that the effect of the relaxation time is more prominent at lower
temperatures (or larger $\alpha_s$).

\begin{figure}[btp]
\epsfxsize=8.5cm
\epsfbox{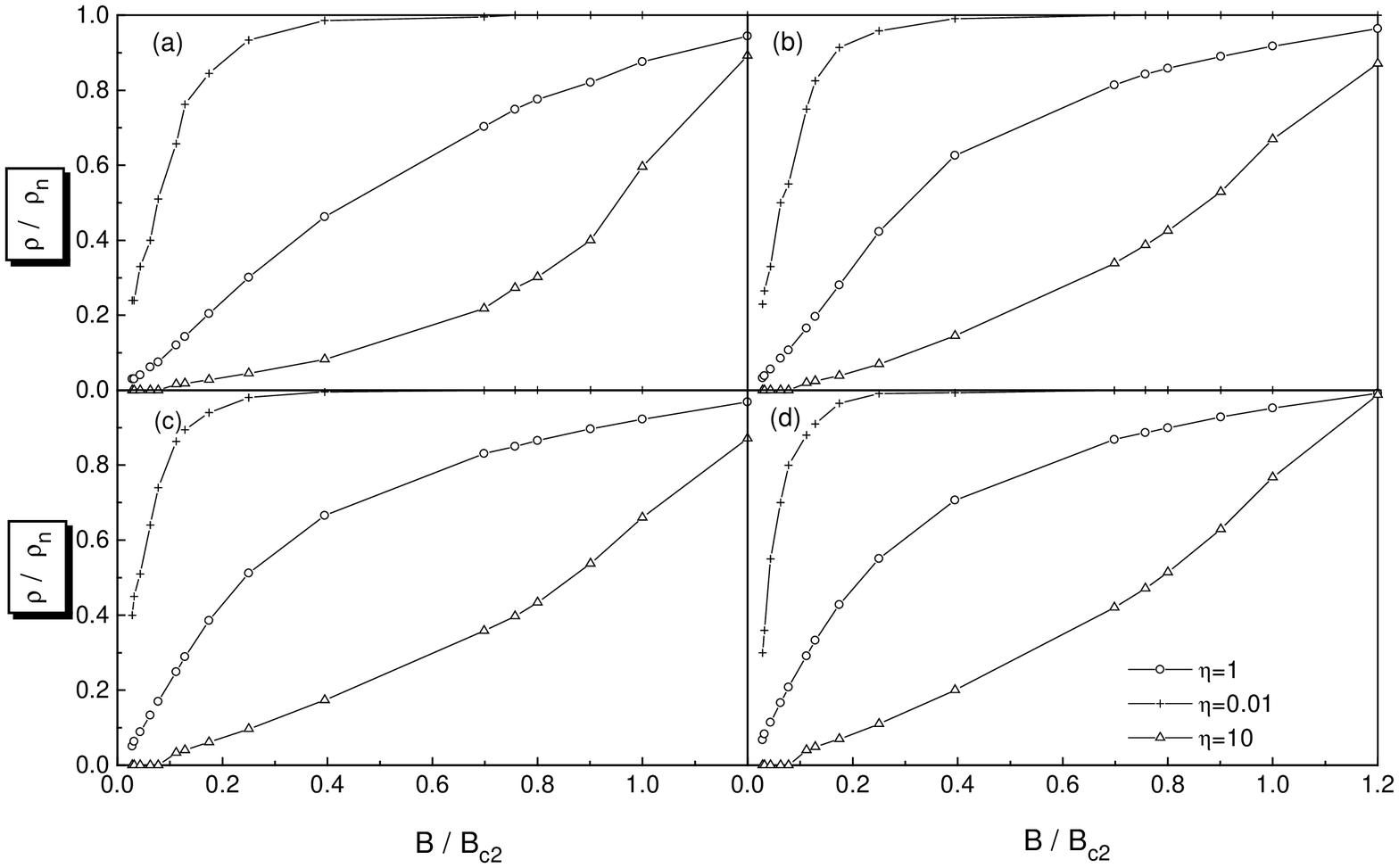}
\caption{Free vortex flow resistivity as a function of the applied field for 
$\eta_s=1$, $10$, and $0.01$. Here $\alpha_{s}=0.97$, $0.85$,
$0.67$, and $-1$ in (a), (b), (c), and (d) respectively.}
\end{figure}

Next, we investigate the vortex motion of the $d+is$ superconductor in the
presence of twin-boundaries and then look into the pinning effect. A
periodic array of twin-boundaries (with a transverse spacing of $L=10.8\xi$)
are assumed and described by $\alpha_{i}=\alpha_{i,0}-
u_{i}\sum\limits_{k}\delta(y-y_{0}-kL)$, where the subscript $i$ stands for $%
s$ or $d$. Here, $u_i$ describes the variation of $\alpha_{s,d}$ across the
twin boundary along the line $y=y_{0}$ due to the local mis-orientation or
chemical contamination. Along the twin-boundary, we apply a transport
current ${\bf J}=J\hat{x}$. The vortex motion will be pinned by the
twin-boundary up to a depinning current $J=J_c$, which we shall determine.
In the following simulation we fix $\eta_{s}=2\eta_{d}=1$. Figure 4 gives
the current dependence of the flux-flow resistivity, which turns out to be
highly nonlinear. In the present simulation, we may expect a simple result
of the overdamped vortex motion in a periodic pinning potential: $%
\rho/\rho_n=a\sqrt{1-(J_{c}/J)^2}$ at $J\geq J_c$(the solid lines in Fig.4),
~\cite{Wang} where $a$ is the asymptotic reduced resistivity and $J_c$ can
be thus determined. We see that a higher depinning current arises from
higher values of $u_s$ (or $u_d$). This is because of the increasing
suppression of the the amplitudes of $s$- wave and $d$- wave components at
the boundary. In fact, with a suppression of the order parameter at the
twin, the vortices loose less condensation energy, and thus have a lower
energy than they would do in the bulk. For a $d$-wave superconductor, $u_s$
is irrelevant because $\alpha_{s}<0$, only $u_d$ takes effect, so it is
reasonable that a $d$- wave superconductor has a lower depinning current
than a $d+is$ superconductor does with the same twin boundary.

\begin{figure}[btp]
\epsfxsize=8.5cm
\epsfbox{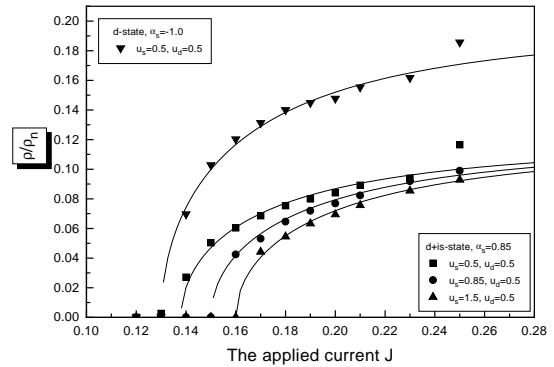}
\caption{The current dependence of the flux-flow resistivity 
in the presence of a twin-boundary. The solid lines 
represent the fitting at low $J$(see the text).}

\end{figure}

Finally, we turn our attention to the so called intrinsic Hall effect, which
occurs for anisotropic vortex structures even without considering normal
Hall conductivity $\sigma_{H}^n$. Alvarez {\it et al.} have found that in $d$%
- wave superconductors the intrinsic Hall effect depends on the orientation
angle, $\varphi$, of the driving current roughly as $\sim sin(4\varphi)$,
and increases nonlinearly with ${\bf J}$.~\cite{Alvar} It is believed that
this effect is due to the four-lobe structure of the $d$- wave vortices. So
here we would like to explore the effect of the two-fold symmetry of $s$-
and $d$- components on the intrinsic Hall effect. As usual, the intrinsic
Hall effect is measured by $\tan \theta_{H} =E_{AV,\perp}/E_{AV,\parallel}$,
where ${\bf E}_{AV}$ and $E_{AV,\perp}$ are the components of ${\bf E}_{AV}$
perpendicular and parallel to the applied current, respectively. Figure 5
presents $\tan \theta_H$ versus $\varphi$ for (a) a $d$- wave superconductor
with $\alpha_s=-1.0$ and $\alpha_d=1.0$; (b) a $d+is$- wave superconductor
with $\alpha_s=0.85$ and $\alpha_d=1.0$. Here $B=0.034B_{c2}$, and $|{\bf J}%
|=0.08$ at which $\bar{\rho}/\rho_n \approx 0.6$ in both cases (a) and (b).
It can be seen that for the $d$- wave superconductor (Fig. 5(a)), $\tan
\theta_H$ has two peaks (at $\varphi_p=15^{\circ}$ and $\varphi_p=75^{\circ}$%
), and is identically zero at $\varphi=0^{\circ}, 45^{\circ}$, and $%
90^{\circ}$. For the $d+is$ superconductor, $\tan \theta_H$ is still zero at 
$\varphi=0^{\circ}, 45^{\circ}$ and $90^{\circ}$, but there are new sign
reversals at approximately $\varphi=22.5^{\circ}$ and $67.5^{\circ}$.
Clearly, the vanishing of $\tan \theta_H$ at $\varphi=0^{\circ}, 45^{\circ}$
and $90^{\circ}$ can be attributed to the fact that the vortex(and thus the
supercurrent around it) can adjust  its symmetry axis to the the direction
of the driving force from the the applied current. At other directions of
the driving force, the vortex can only partially adjust its symmetry axis.
The new sign reversal for the $d+is$ superconductors is more subtle. In this
case, the vortex profile is two-fold symmetric. Therefore, there are two
non-equivalent configurations for the vortex to have its (reflection)
symmetry axis along the $x$, $y$, or the diagonal directions. We suspect
that the abnormal sign reversal results from a switching between these two
meta-stable configurations as the direction of the driving force changes.

\begin{figure}[btp]
\epsfxsize=8.5cm
\epsfbox{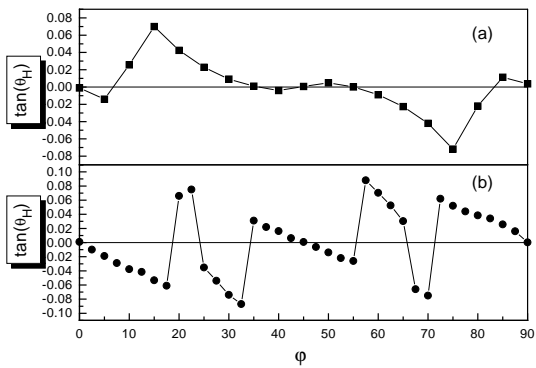}
\caption{$\tan \theta_H$ as function of the current orientation
angle $\varphi$ at $|{\bf J}|=0.08$ and $B=0.034B_{c2}$. 
(a) $\alpha_s=-1.0$ and $\alpha_d=1.0$; (b) $\alpha_s=0.85$, 
$\alpha_d=1.0$. The lines are guides for the eye.}

\end{figure}

In summary, the dynamics of the vortices in both $d+is$-wave and $d$-wave
superconductors is studied. The upper critical field of the $d+is$-wave
superconductors is studied analytically and numerically. From simulation
results, the curvature of the FFF resistivity as a function of the magnetic
field strongly depends on the relaxation rate of the order parameter. The
flux flow in the presence of a twin boundary is also addressed. Finally, the
intrinsic Hall effect of $d$- and $d+is$- wave superconductors are studied
numerically . We find the orientation dependence of the intrinsic Hall
effect in these two types of superconductors are rather different.

This work was supported by the RGC grant of Hong Kong under No. HKU 7116/98P
and a CRCG grant at the University of Hong Kong. Q.H.W. was supported by the
National Natural Science Foundation of China and in part by the Ke-Li
Fellowship financed by Sanzhu Co. Ltd. in Shandong of China.


\begin{references}
\bibitem{Note}  The physical properties as a function of temperature
discussed later  are only qualitative, as the present $T$-dependence of $%
\alpha_{s}$  is not quantitatively rigorous in the whole temperature range.
\end{references}
\end{document}